\documentclass{Interspeech}

\interspeechcameraready

\title{SC-SOT: Conditioning the Decoder on Diarized Speaker Information\\for End-to-End Overlapped Speech Recognition}

\author[affiliation={1}]{Yuta}{Hirano}
\author[affiliation={1}]{Sakriani}{Sakti}

\affiliation{}{Nara Institute of Science and Technology}{Japan}
\email{yuta.hirano.ia4@naist.ac.jp, ssakti@is.naist.jp}

\usepackage{comment}
\newcommand{\1}{\mbox{1}\hspace{-0.25em}\mbox{l}}

\begin{document}

\maketitle

\begin{abstract}

    We propose \textbf{S}peaker-\textbf{C}onditioned \textbf{S}erialized \textbf{O}utput \textbf{T}raining (SC-SOT), an enhanced SOT-based training for E2E multi-talker ASR. 
    We first probe how SOT handles overlapped speech, and we found the decoder performs implicit speaker separation. 
    We hypothesize this implicit separation is often insufficient due to ambiguous acoustic cues in overlapping regions. To address this, SC-SOT explicitly conditions the decoder on speaker information, providing detailed information about ``who spoke when''. Specifically, we enhance the decoder by incorporating: (1) speaker embeddings, which allow the model to focus on the acoustic characteristics of the target speaker, and (2) speaker activity information, which guides the model to suppress non-target speakers. The speaker embeddings are derived from a jointly trained E2E speaker  diarization model, mitigating the need for speaker enrollment.
    Experimental results demonstrate the effectiveness of our conditioning approach on overlapped speech. 

\end{abstract}

\section{Introduction}
Since the development of the first speech recognition system over half a century ago, speech recognition technology has steadily advanced. In particular, recent advances in automatic speech recognition (ASR) have been remarkable, largely due to the rise of deep learning \cite{li2022recent, prabhavalkar2023end}. In common single-speaker tasks, especially, models surpassing human parity are being reported one after another \cite{radford2023robust, ramirez2024anatomy, puvvada2024less}. 

Given such impressive achievements, ASR may seem like a solved problem. However, there are still many challenges to be addressed. One of the most significant challenges is overlapped speech recognition, which is referred as multi-talker ASR. Multi-talker ASR deals with the scenario where multiple speakers are talking simultaneously, resulting in overlapped speech segments. This presents a significant departure from the well-studied single-talker ASR scenario, where the input audio segment is assumed to contain only one speaker at a time. The core challenge in multi-talker ASR lies in disentangling the mixed acoustic signals to accurately transcribe the speech of each individual speaker. 
The ability to accurately process and transcribe multi-talker speech is crucial for a wide range of applications, including automated meeting transcription and analysis of multi-party conversations. 

Driven by the rapid development of deep learning, recent research has increasingly focused on end-to-end (E2E) models for multi-talker ASR. These E2E models offer the potential for joint optimization of all components through back-propagation, leading to improved performance and simplified system design. A number of promising E2E approaches have emerged \cite{yu2017recognizing, kanda2020serialized, shi2020sequence, kanda2022streaming, raj2023surt}. The initial approach was proposed by Yu \textit{et al.} \cite{yu2017recognizing}. In this work, the permutation invariant training (PIT) was applied to a neural network with two output channels, enabling the transcription of overlapped speech. PIT was originally proposed for speech separation \cite{yu2017permutation} and was later applied to E2E multi-talker ASR \cite{yu2017recognizing}. However, a major drawback of PIT is that the number of speakers that can be handled is limited to a predetermined number. In fact, most PIT-based methods could only handle two speakers \cite{yu2017recognizing, yu2017permutation, seki2018purely, chang2019mimo}. To address this limitation, serialized output training (SOT) \cite{kanda2020serialized} was proposed. In SOT, unlike PIT, transcriptions for all speakers is predicted as a single token sequence. As a result, SOT does not require a predetermined number of speakers and can theoretically handle arbitrary number of speakers.

Although SOT successfully addresses the permutation problem and allows for a variable number of speakers, handling heavily overlapped speech is still a significant challenge for SOT. This may be because SOT-based models basically learn how to transcribe overlapped speech only from labels serialized in speakers' speaking start times. To overcome this limitation, we propose to enhance SOT by explicitly providing the model with information about ``who spoke when''. 
This is achieved through our proposed approach called \textbf{S}peaker-\textbf{C}onditioned \textbf{S}erialized \textbf{O}utput \textbf{T}raining (SC-SOT), which conditions the decoder on speaker information. Specifically, SC-SOT incorporates two key components: (1) speaker embedding, derived from a jointly trained E2E neural speaker diarization with encoder-decoder based attractors (EEND-EDA) model \cite{horiguchi2020end}, 
which allows the model to focus on the acoustic characteristics of the current target speaker, and (2) speaker activity information, also obtained from the EEND-EDA, which helps suppress non-target speakers or silence. By incorporating this explicit speaker information, SC-SOT aims to enhance the implicit speaker separation performed by the decoder. Moreover, our proposed approach mitigates the need for speaker enrollment by leveraging the jointly trained EEND-EDA.
The main contributions of this paper can be summarized as follows.
\begin{itemize}
\item We show that the decoder plays a crucial role in handling overlapped speech by visualizing attention weights.
\item We propose a conditioning method for the decoder using speaker embeddings and speaker activity information without the need for speaker enrollment.
\item We propose a method to integrate speaker counting by the diarization module into SOT-based E2E multi-talker ASR.
\item Experimental results demonstrate our proposed approach improves SOT-based E2E multi-talker ASR in terms of both speech recognition accuracy and speaker counting accuracy.
\end{itemize}

\section{Related Works}
\subsection{Handling overlapped speech using speaker embeddings}
Target speech extraction is a task that aims to isolate the speech of a specific speaker from a mixture containing multiple speakers, relying on cues about the target speaker's identity. Speaker embedding is a common type of information used in target speech extraction \cite{zmolikova2023neural, vzmolikova2019speakerbeam, jansky2020adaptive}. Target-speaker voice activity detection (TS-VAD) \cite{medennikov2020target, wang2023target} is another successful approach that leverages speaker embeddings to handle overlapped speech. TS-VAD is one of a mainstream technique in speaker diarization, solving the diarization problem by performing voice activity detection for each speaker, conditioned on their respective speaker embeddings. These studies suggest that using speaker embeddings as conditioning information is effective for handling overlapped speech.

Beyond speech separation and speaker diarization tasks, speaker embeddings have also been integrated into E2E multi-talker ASR. 
Huang \textit{et al.} \cite{huang2023adapting} proposed an E2E multi-talker ASR model based on self-supervised learning (SSL) models, using speaker embeddings. They demonstrated the effectiveness of speaker embeddings for multi-talker ASR. However, the number of speakers their model can handle is fixed due to the use of connectionist temporal classification (CTC) decoder \cite{graves2006connectionist}, which typically requires a fixed output dimension corresponding to the number of speakers. Furthermore, this approach assumes the availability of speaker enrollment data for each speaker during both training and inference. The requirement for speaker enrollment poses a significant limitation in many real-world applications. 
 
\subsection{Handling overlapped speech using speaker activity information}
Relatively few studies have explored the use of speaker activity information for handling overlapped speech, especially when compared to the extensive work on speaker embedding-based methods.
Polok \textit{et al.} \cite{polok2024target, polok2024dicow} proposed a conditioning method to adapt Whisper \cite{radford2023robust} for target-speaker ASR, utilizing speaker activity information instead of speaker embeddings 
to extract the target speaker's representation from intermediate encoder layers. This approach achieves target-speaker ASR without requiring the learning of a complete speaker embedding space. 
However, relying solely on speaker activity information is often insufficient for accurately handling overlapped speech, particularly when the overlapping speakers exhibit similar activity patterns within input speech segments.
Our proposed approach addresses this limitation by leveraging not only speaker activity information but also the corresponding speaker embeddings. This serves as a more robust and discriminative representation of each speaker, even in challenging overlapping conditions.

\begin{figure}
    \centering
    \includegraphics[width=0.9\linewidth]{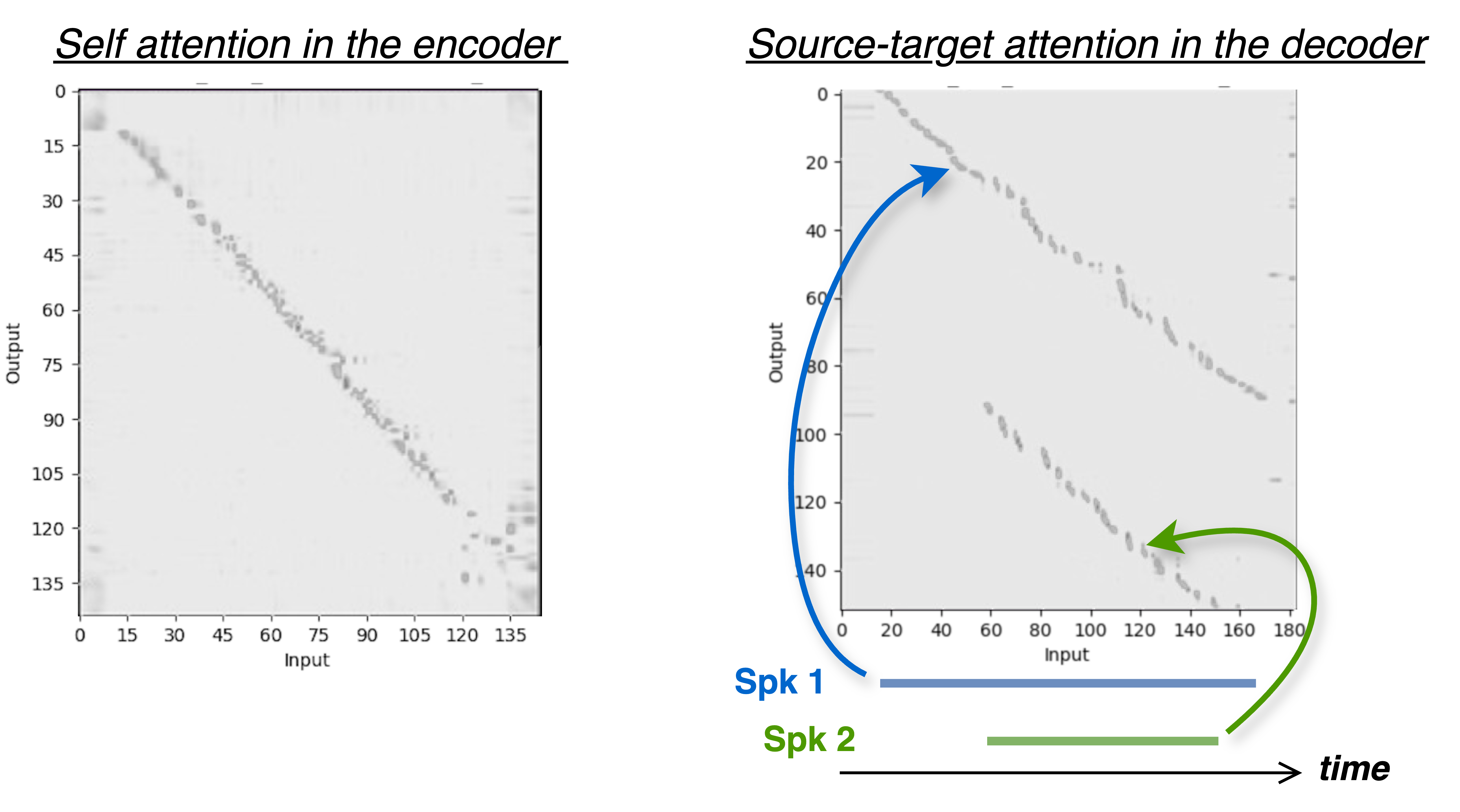}
    \caption{The visualized attention weights of a SOT-based model in two-speaker overlap case. (Left) Self attention in the encoder and (Right) source-target attention in the decoder. In the visualized source-target attention, the attended frames in each diagonal line align approximately with the speaker's active time.}
    \label{fig:enter-label}
\end{figure}

\section{Proposed Approach}

\begin{figure*}[t]
    \centering
    \includegraphics[width=1\linewidth]{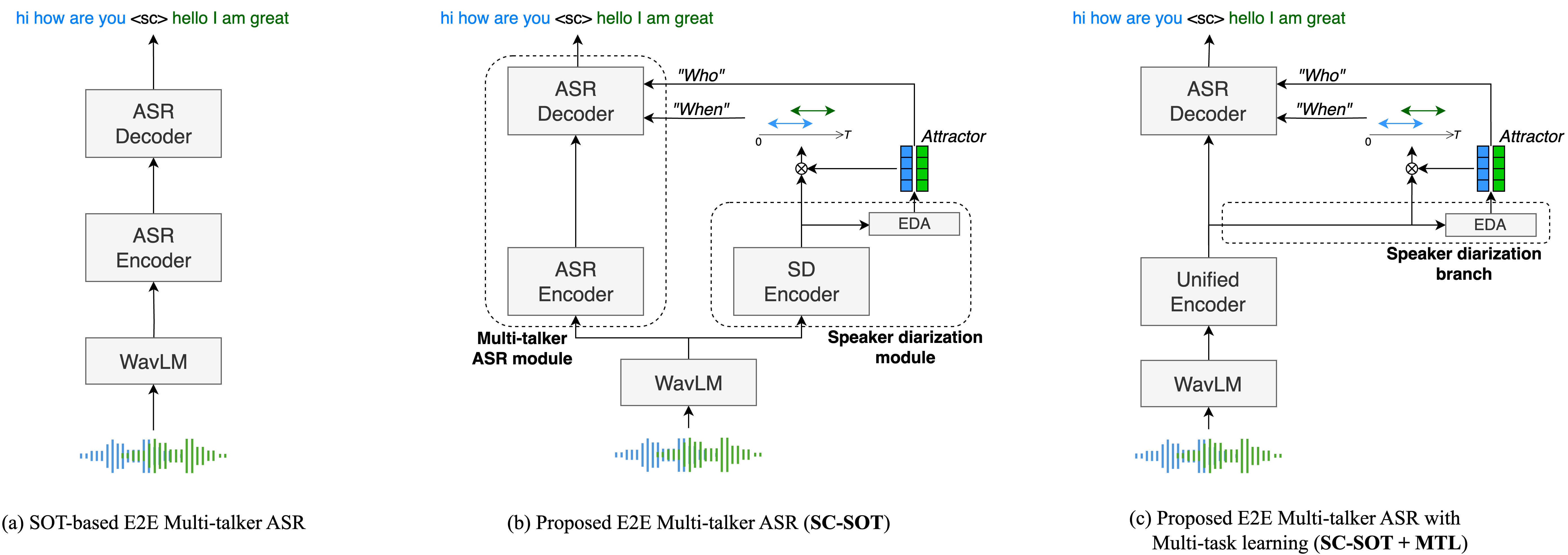}
    \caption{Overview of the conventional SOT-based end-to-end multi-talker ASR and the proposed approach. }
    \label{fig:overall_model_comparison}
\end{figure*}

\subsection{Encoder and decoder behavior in handling overlapped speech}
Although SOT-based models can transcribe overlapped speech without separating it, how SOT-based models achieve this has not been revealed. To understand SOT-based models' behavior, we explored attention weights in the model. 
Figure \ref{fig:enter-label} shows the examples of visualized attention weights of both the encoder and the decoder in an SOT-based model. Looking at the source-target attention weight, it turned out that the relationship between the target token sequence and the encoder representation is not fully monotonic and overlapped segments were attended twice. Moreover, the encoder maintains temporal relationship between the frames and does not perform any separation. These findings indicate that the encoder keeps the temporal information of input overlapped speech and the decoder performs speaker separation implicitly. This finding greatly influenced the design of the proposed approach.

\subsection{Conditioning the decoder on speaker information}
Serialized output training (SOT) \cite{kanda2020serialized} is a simple training strategy for E2E multi-talker ASR. Figure \ref{fig:overall_model_comparison} (a) shows the overview of SOT-based multi-talker ASR model. 
SOT does not require a specialized model architecture to handle overlapped speech; the same model used for single-talker scenarios can be employed. However, the applicable models are generally limited to attention-based encoder-decoder architectures. This restriction is due to the fact that the relationship between input speech and the target token sequence is not strictly monotonic in SOT. SOT models multi-talker ASR as follows. First, SOT introduces a special token $\langle{sc}\rangle$ that represents speaker change. Second, the target sequence is created by serializing multiple references from different speakers into a single sequence and inserting $\langle{sc}\rangle$ between them. For example, for a two-speaker case, the target sequence is $\{y_1^1, \cdots, y_{N_1}^1, \langle{sc}\rangle, y_{1}^2, \cdots, y_{N_2}^2, \langle{eos}\rangle\}$, where $N_1$ and $N_2$ are the lengths of the references from the first and second speakers, respectively. References from multiple speakers are usually serialized in first-in first-out (FIFO) order based on their speaking start time.

We propose to enhance SOT by conditioning the decoder on speaker embeddings and speaker activity information. To obtain those two information from input overlapped speech, SC-SOT has a speaker diarization module (branch) based on E2E neural diarization with encoder-decoder based attractors (EEND-EDA) \cite{horiguchi2020end}. This speaker diarization module can predict speaker activities and local speaker embeddings called attractor from overlapped speech as well as EEND-EDA. This speaker diarization module is jointly trained with multi-talker ASR module, and the FIFO speaker order is used for the label permutation of the speaker diarization labels instead of the permutation given by PIT \cite{yu2017permutation}. SC-SOT-based model is optimized using the following loss function:
\begin{equation}
\label{loss_func}
    \mathcal{L} = \mathcal{L}_{asr} + \alpha\mathcal{L}_{diar},
\end{equation}
where $\mathcal{L}_{asr}$ and $\mathcal{L}_{diar}$ are the loss function of SOT and EEND-EDA, respectively. $\alpha$ is a hyperparameter to adjust the loss for speaker diarization task.
Figure \ref{fig:overall_model_comparison} (b) and (c) provide the overview of the proposed SC-SOT with and without multi-task learning.

\subsubsection{Conditioning the decoder on speaker embeddings}
The speaker embeddings predicted by the speaker diarization block are used to allow the model to focus on the acoustic characteristics of the target speaker.
To condition the decoder on the speaker embeddings, the position-wise feed-forward network (FFN) in the $l$-th transformer decoder layer is modified as follows:
\begin{align}
        \text{FFN}_l (\mathbf{z}_{n-1,s}+\mathbf{w}^{spk} \cdot \mathbf{a}_s) \quad(l=1) \nonumber\\
        \text{FFN}_l (\mathbf{z}_{n-1,s}) \quad(l>1),
\end{align}
where $\mathbf{z}_{n-1,s}$ is the decoder representation calculated from $n-1$th token for the $s$-th speaker, $\mathbf{a}_s$ is the attractor of the $s$-th speaker, and $\mathbf{w}^{spk}$ is a learnable weight matrix. 
This formulation shares an idea with Transcribe-to-Diarize \cite{9746225}. However, the objectives differ, and several aspects of our approach are also distinct: the method for acquiring speaker embeddings, multi-task learning with E2E speaker diarization, 
and functions described in remaining sections.

\subsubsection{Conditioning the decoder on speaker activity information}
The speaker activity information predicted by the speaker diarization block is used to guide the model to suppress non-target segments. We selectively apply attention penalty to source-target attention in the decoder. This method is originally proposed by Polok \textit{et al.} \cite{polok2024but} to convert Whisper into a target speaker ASR model. We adopt this method for SOT-based E2E multi-talker ASR.
In each attention head of the source-target attention, we penalize the source-target attention scores in non-target speaker segments by modifying them as follows:
\begin{align}
\label{activity1}
    Penalized\_Attention = \mathbf{QK}^{T} - \mathbf{c},
\end{align}
where the penalty, $\mathbf{c}$=$\{c_{1}, c_{2}, \dots,c_{T}\}$, is calculated as:
\begin{align}
\label{activity2}
    c_{t} = c \odot \1(p_{t, s}^{diar} < \theta),
\end{align}
where $\mathbf{QK}^{T}$ is the original source-target attention score before normalization, $c$ is a hyperparameter controlling the strength of the penalty, $p_{t, s}^{diar}$ represents the posterior probability for speaker $s$ at time frame $t$ predicted by the speaker diarization module, and $\theta$ is a threshold determining whether a speaker is considered active.  $\odot$ and $\1$ denote the Hadamard product and the indicator function, respectively.

\subsubsection{Speaker counting by EEND-EDA}
In addition to the two conditioning methods described above, we propose incorporating speaker counting by EDA module into decoding. 
In conventional SOT-based models, decoding terminates when the end-of-sequence ($\langle{eos}\rangle$) token is output. 
While SOT-based models have been shown to have high speaker counting accuracy, it is still not perfect. Furthermore, because the speaker counting task is not explicitly trained, there may be room for improvement in speaker counting through supervised learning. EEND-EDA optimizes speaker counting through discriminative training. Therefore, enforcing the number of speakers predicted by the speaker diarization module to the decoding may improve speaker counting of SOT-based models.
To enforce the number of speakers predicted by the speaker diarization module during decoding, we made the following modifications. First, we replaced the $\langle{eos}\rangle$ token at the end of the label with the $\langle{sc}\rangle$ token. This prevents the decoding process from stopping automatically. Second, decoding is terminated when the number of generated $\langle{sc}\rangle$ tokens equals the number of speakers predicted by the speaker diarization module.

\section{Experimental Setup}

\begin{table*}[h]
    \centering
    \caption{WER, SCA and DER (dev / test) of our proposed SC-SOT with different conditioning configurations compared to the conventional SOT on Libri2Mix and Libri3Mix. MTL denotes multitask learning with the ASR objective and the speaker diarization objective. ``*'' indicates the use of oracle speaker diarization information.}
    \label{tab:results_librimix}
    \begin{tabular*}{\textwidth}{ll@{\extracolsep{\fill}}cccccccc} \hline\hline
         & &  \multicolumn{2}{c}{\textbf{Information to condition}} & \multicolumn{3}{c}{\textbf{Libri2Mix}} & \multicolumn{3}{c}{\textbf{Libri3Mix}} \\
         \cmidrule(lr){3-4} \cmidrule(lr){5-7} \cmidrule(lr){8-10}
        &\textbf{Model} & {\textbf{\begin{tabular}[c]{@{}c@{}}Speaker\\Emb.\end{tabular}}} & {\textbf{\begin{tabular}[c]{@{}c@{}}Speaker\\Activity Info.\end{tabular}}} & {\textbf{WER(\%)}} & {\textbf{SCA(\%)}} & {\textbf{DER(\%)}} & {\textbf{WER(\%)}} & {\textbf{SCA(\%)}} & {\textbf{DER(\%)}} \\ \hline
        1&SOT \cite{watanabe2018espnet} & - & - & 19.7 / 17.1 & - & - & - & - & - \\
        2&SOT (ours) & - & - & 19.2 / 16.8 & 98.8 / 99.0 & - & 36.0 / 34.5 & 89.2 / 88.1 & - \\ \hline
        3&SOT + MTL & - & - & 19.8 / 17.7 & 99.6 / 99.5 & 3.6 / 3.7 & 34.5 / 32.9 & 93.4 / 93.4 &  8.2 / 8.1\\ 
        4&SC-SOT & \checkmark & $\times$ & 18.0 / 15.6 & 99.6 / 99.4 & 4.6 / 4.3&34.3 / 33.1  &88.4 / 86.2  & 21.9 / 22.1 \\
        5&SC-SOT + MTL & \checkmark & $\times$ & \textbf{17.5} / \textbf{15.2}& \textbf{99.9} / \textbf{99.9}& 8.2 / 8.9 &\textbf{33.2} / \textbf{30.9}  &95.4 / 95.2  & 22.6 / 23.4 \\
        6&SC-SOT + MTL  & $\times$ & \checkmark & 17.8 / 15.7 & 99.8 / \textbf{99.9} & 3.1 / 3.1 & 34.5 / 32.5&93.7 / 93.9  &7.2 / 7.4 \\
        7&SC-SOT + MTL & \checkmark & \checkmark & 18.7 / 16.2& 99.2 / 99.5& 7.0 / 7.3&36.2 / 34.6  & \textbf{96.0} / \textbf{95.3} & 20.6 / 21.2\\ \hline
        8&SC-SOT + MTL & \checkmark & \checkmark* & 17.4 / 15.1& - & - & 31.1 / 29.2& - & - \\ \hline\hline
    \end{tabular*}
\end{table*}

\subsection{Dataset}
The training set consists of LibriSpeech-360h \cite{panayotov2015librispeech}, Libri2Mix, and Libri3Mix \cite{cosentino2020librimix}.
Libri2Mix and Libri3Mix were modified by introducing utterance start time offsets. For Libri2Mix, the simulating scripts and metadata provided by ESPnet \cite{watanabe2018espnet} were used. The simulating scripts and metadata for generating the training set from Libri3Mix were created by following the same simulation protocol as Libri2Mix.
We used the test-set of Libri2Mix and Libri3Mix for evaluation.

\subsection{Model configuration}
The baseline SOT-based model employs a Conformer encoder \cite{gulati2020conformer} and a Transformer decoder \cite{vaswani2017attention}. The Conformer encoder has 12 layers, while the Transformer decoder is structured with 6 layers. Both the encoder and decoder utilize 4 attention heads and a hidden dimension of 256 and a 2048-dimensional FFN. The kernel size of the convolution layer in the Conformer encoder was set to 31. Input features were derived from the WavLM large \cite{chen2022wavlm} representation, using a weighted sum with a learnable weight. 
The multi-talker ASR module of all the proposed model employs the same architecture as the baseline model. When an independent speaker diarization module is used, its encoder is built with 4 layers of Conformer, incorporating 4 attention heads, a hidden dimension of 256, and a 1024-dimensional FFN. 
The EDA module consists of one unidirectional LSTM layer with 256 hidden units. During training, the ground truth speaker diarization label were used for conditioning. The hyperparameter $\alpha$ in equation \ref{loss_func} was set to 0.1. The hyperparameter $c$ and $\theta$ in equation \ref{activity1} and \ref{activity2} were set to 50 and 0.5, respectively.
All models were trained using the Adam optimizer \cite{kingma2014adam}. The learning rate was set to 0.001 and warmed up for 10000 iterations. 
The batch bin size was set to $2.4\text{e}7$ and trained all models for 50 epochs.
Character was used as the output unit. 
Word error rate (WER) and speaker counting accuracy (SCA) were used for evaluation matrices. We report diarization error rate (DER) with collar tolerance of 0.0 sec and median filtering of 11 frames as a supplemental result.

\section{Experimental Results}

Table \ref{tab:results_librimix} presents the performance of our proposed SC-SOT-based model on the Libri2Mix and Libri3Mix datasets, alongside baseline comparisons and variations in conditioning configurations. 
Our proposed SC-SOT models, particularly when combined with multi-task learning (MTL), demonstrate promising performance gains over the conventional SOT-based model. On Libri2Mix, the best performing SC-SOT+MTL model (using speaker embeddings only) achieves a WER of 17.5\% / 15.2\%, a substantial improvement over the baseline SOT's 19.2\% / 16.8\%. Similarly, on Libri3Mix, the same SC-SOT+MTL configuration yields a WER of 33.2\% / 30.9\%, outperforming the baseline SOT. Introducing speaker embeddings alone (rows 5 and 6) consistently lowers WER compared to unconditioned SOT models, highlighting the utility of speaker identity information for improved implicit speaker separation by the decoder. 
It is important to note that these WER improvements were achieved without any additional inputs.

Incorporating only speaker activity information (row 6) also contributes to WER reduction. However, the combination of conditioning on both speaker embeddings and speaker activity information (row 7) showed WER increase from SC-SOT with a single condition. Particularly, SC-SOT earned worse WER than SOT on Libri3Mix. This WER increase seems to be caused by many errors in speaker activity information. We consistently observed that the accuracy of the diarization module is distorted when speaker embeddings are used for conditioning during training (row 4, 5, and 7). This may indicate the optimal representation of speaker embeddings is different between ASR module and diarization module, and the gradient from SOT objective guided the representation of speaker embeddings to be more optimal for ASR task than for diarization task.

To isolate the potential of combining both conditioning inputs, we conducted an evaluation using oracle speaker diarization information alongside the predicted speaker embeddings (row 8). This configuration reduced WER, particularly on Libri3Mix. This result strongly indicates that the concept of combining speaker embeddings and activity information is highly beneficial, and that the performance limitations observed in row 7 are primarily due to the accuracy of the speaker diarization module, rather than an inherent incompatibility between the two conditioning types. The gap between the performance with predicted (row 7) and oracle (row 8) speaker activity information highlights the substantial gains achievable with improved speaker diarization accuracy.

\section{Conclusion}

This work investigated and enhanced the implicit speaker separation capabilities of SOT-based models. Our initial analysis, through visualization of source-target attention weights in an SOT-based model, revealed that the decoder performs a degree of speaker separation. Building upon this key insight, we proposed an approach: conditioning the decoder on speaker embeddings and speaker activity information to explicitly augment this implicit separation process. 
We achieved relative reductions in WER of up to 9.5\% on Libri2Mix and 10.4\% on Libri3Mix. Furthermore, experiments utilizing oracle speaker diarization information as speaker activity input yielded even greater accuracy. This highlights the significant potential of our approach and points towards future research directions focused on refining the speaker diarization component.

\clearpage
\section{Acknowledgements}
Part of this work is supported by JSPS KAKENHI Grant Numbers JP21H05054 and JP23K21681.
\bibliographystyle{IEEEtran}
\bibliography{template}

\end{document}